%% file: main.tex
\documentclass{article}

\usepackage[preprint]{neurips_2026_vericode}

\usepackage[utf8]{inputenc} 
\usepackage[T1]{fontenc}    

\usepackage[dvipsnames]{xcolor}
\usepackage{hyperref}
\hypersetup{hypertexnames=false,colorlinks=true,linkcolor=NavyBlue,citecolor=NavyBlue,urlcolor=ForestGreen}

\usepackage{tikz}
\usetikzlibrary{arrows.meta,positioning,fit,calc}

\usepackage{url}            
\usepackage{booktabs}       
\usepackage{amsfonts}       
\usepackage{nicefrac}       
\usepackage{microtype}      
\usepackage{xcolor}         
\usepackage{graphicx}
\usepackage{algorithm}
\usepackage{algorithmic}
\usepackage{amsthm}
\usepackage{amsmath}
\usepackage{amssymb}
\usepackage{mathtools}
\usepackage{soul}
\usepackage{listings}
\newcommand{\change}[1]{{\color{black}{#1}}}
\lstdefinestyle{gallerylean}{
  basicstyle=\ttfamily\small, columns=fullflexible,
  keepspaces=true, showstringspaces=false, breaklines=true,
  frame=single, rulecolor=\color{NavyBlue!40},
  backgroundcolor=\color{NavyBlue!3}, framesep=5pt,
  aboveskip=5pt, belowskip=8pt,
  literate={→}{{$\rightarrow$}}1 {∀}{{$\forall$}}1
    {∃}{{$\exists$}}1 {∧}{{$\land$}}1 {↔}{{$\leftrightarrow$}}1
    {≤}{{$\leq$}}1 {≥}{{$\geq$}}1 {≠}{{$\neq$}}1
    {∈}{{$\in$}}1 {·}{{$\cdot$}}1
}
\newcommand{\galleryproblem}[1]{\par\smallskip\noindent
  \fcolorbox{NavyBlue!40}{NavyBlue!3}{%
    \begin{minipage}{\dimexpr\linewidth-2\fboxsep-2\fboxrule\relax}
    \small\textbf{Natural-language problem (source excerpt).}\par\smallskip #1
    \end{minipage}}\par\medskip}

\theoremstyle{plain}

\theoremstyle{definition}

\theoremstyle{remark}

\renewcommand{\cite}{\citep}

\bibpunct{(}{)}{;}{a}{,}{,}

\title{Beyond Type-checking: Towards Holistic Evaluation of Formal Specification
Generation}

\author{%
    Srijith Nair\thanks{Work done as a summer AI intern at Emergence AI.} \\
    Department of Electrical \& Computer Engineering\\
    The Ohio State University\\
    \texttt{nair.203@osu.edu} \\
    \And
    Aditya Vempaty \\
    Emergence AI \\
    \texttt{aditya@emergence.ai} \\
    \And
    Jia Liu \\
    Department of Electrical \& Computer Engineering \\
    The Ohio State University\\
    \texttt{liu@ece.osu.edu} \\
    \And
    Ashish Jagmohan \\
    Emergence AI \\
    \texttt{ashish@emergence.ai}
}

\begin{document}

\maketitle

\begin{abstract}
    \input{sections/abstract}
\end{abstract}

\section{Introduction}
\input{sections/intro}

\section{Benchmark Design}
\input{sections/design}

\section{Experimental Results}
\input{sections/results}

\section{Related Work, Limitations, and Future Work}
\input{sections/future}

\section{Conclusion}
\input{sections/conclusion}

\clearpage
\bibliographystyle{plainnat}
\bibliography{references}
\clearpage
\appendix
\input{sections/appendix_examples}

\input{sections/appendix_audits}


\end{document}

%% file: sections/abstract.tex
When generating verifiable code, natural language requirements are mapped to
machine checked code using LLMs and agentic workflows.
A crucial component of this pipeline is specification generation (SpecGen),
which produces a formal contract against which an agent can prove implementation
correctness.
Proof generation can obtain deterministic feedback from a theorem prover, but
SpecGen lacks a definitive check that a generated specification captures the
user's intent.
A checked proof can therefore establish correctness against a specification that
misrepresents the intended behaviour.
We take a step towards holistic SpecGen evaluation with a unified dataset
assembled from 350 existing Lean tasks, including 189 from VERINA and 161 from
CLEVER, and a framework covering formal validity, reference similarity and
equivalence, and behavioural adequacy.
We distinguish acceptance of required inputs from acceptance of valid outputs
and rejection of invalid outputs, while making each metric's evidence scope
explicit.
Across four SpecGen configurations, restricting the \emph{generalized tree edit
distance} (GTED) comparison, a reference similarity measure, to 32 jointly
measurable VERINA tasks changes the VERINA configuration's position from second
to fourth in mean similarity, showing the importance of measurement coverage.
In an authored control, a specification achieves $100\%$ positive test recall
and negative test rejection while accepting $0\%$ of required inputs.
This demonstrates that perfect postcondition scores can miss an unusable input
contract, motivating separate feedback on input coverage and output constraints.

%% file: sections/intro.tex
With increasing capabilities of LLMs and agentic workflows, automatic generation
of code based on user defined specifications is rapidly becoming a reality.
State-of-the-art agentic coding systems like Codex \cite{Chen:21:preprint:Codex}
and Claude Code \cite{ClaudeCode:26}, can generate code in a specified
programming language due to training and harness engineering that ground the
LLM's output via planning, breaking the implementation into several manageable
pieces and designing test-cases.
While impressive, these systems are still limited fundamentally due to their
inherently probabilistic nature, and the lack of formal correctness guarantees.
Thanks to decades of research in formal methods, \emph{verifiable code
generation} (VCG) has emerged as a promising approach to generating code jointly
with formal proofs of correctness.
Verifying code correctness against a formally specified contract is a critical 
component of the development pipeline in domains where mistakes in code can have
severe consequences, such as safety critical systems in airplanes
\cite{Cousot:05:ESP:Astree,Delmas:07:ISAS:Astree}, e-commerce and
cloud computing applications \cite{Newcombe:15:ACM:AWS}, kernel
\cite{Konnov:19:ACM:tlaplus} and compiler \cite{Leroy:09:ACM} development,
blockchain \cite{Cassez:22:ICTACAS:Ethereum}, defense systems
and space applications \cite{Verhulst:11:Springer}, among many others.

In the pre-LLM era, formal verification was a laborious and expensive process,
requiring highly skilled engineers to manually write formal specifications and
proofs.
Due to the advent of LLMs, formal verification is now being automated, with LLMs
generating formal specifications.
This often involves a multi-step process of generating: 1) a formal specification
contract from a user's requirement, specified in natural language called
\emph{SpecGen}, and 2) a code implementation that satisfies the specification
called \emph{CodeGen}, and 3) obtaining a formal proof of correctness for the
implementation against the specification called \emph{ProofGen}.
A checked proof establishes correctness relative to the formal contract and
its assumptions. Assurance about user intent also requires the contract to
capture the requirement correctly.

Formal correctness claims can be checked by tools such as
Lean 4 \cite{Moura:21:ICAD:Lean4}, Dafny \cite{Leino:10:ICLPAIR:Dafny}, Verus
\cite{Lattuada:23:ACM:Verus}, Rocq \cite{Barras:97:Inria:Coq}, etc.
These tools check formal correctness claims using proof checking or automated
verification.
Generated specifications are mathematically expressed using Hoare logic
\cite{Hoare:69:ACM,Floyd:93:Springer} as follows:
Given a pure function $C$ which maps inputs $x$ to outputs $y$ such that $y =
C(x)$, if $x$ satisfies the \emph{preconditions} $P(x)$, then $y$ must satisfy
the \emph{postcondition} $Q(x, y)$, i.e.,
\begin{equation}
    \forall x, y \text{ s.t. } P(x), y = C(x) \implies Q(x, y).
    \label{eq:hoare}
\end{equation}

\paragraph{Motivation.}
Proof and code generation can be evaluated deterministically by
using feedback from a theorem prover's compilation process once the user's
intent is captured correctly by the formal specification, and
agentic systems like \citet{Agarwal:26:preprint:IDS} are rapidly becoming better
at proving specifications.
While ProofGen in itself can still be a complicated task, it is nevertheless,
more well-defined than SpecGen, since the latter lacks a definitive feedback
mechanism in spite of being a critical component of the VCG pipeline.
In SpecGen, the specification must capture a user's requirements before a proof
of implementation correctness can provide the intended assurance.

Evaluating this \emph{intent-specification gap} is a challenging problem,
primarily because existing metrics for SpecGen evaluation are limited.
For instance, compiler-assisted type checking detects syntax and type errors
in a specification, but does not establish agreement with user intent.
A well-typed or syntactically correct predicate can still express the wrong
relation.
Thus, an agent can cheat the evaluation using type-checking.
Nevertheless, specification-generation pipelines commonly expose compact signals
such as compilation, reference similarity, test success, or equivalence-search
success.
These signals answer different questions, and their failure modes differ as
well.
We take a step toward comprehensive SpecGen evaluation by unifying existing
evaluation signals and asking: \emph{which dimensions of specification quality
do different evaluation signals reveal, and what feedback can they provide to a
SpecGen agent?}
This enables the development of a high-quality agentic SpecGen system that
can leverage this feedback in the right manner.

\paragraph{Contributions.}
Existing benchmarks already recognize specification generation as a distinct
task in the VCG pipeline.
VERINA \cite{Ye:25:AI4Math:Verina} supports modular code, specification, and
proof evaluation; CLEVER \cite{Thakur:26:NIPS:Clever} evaluates specification
certification relative to a hidden reference.
Our contributions add to these benchmarks richer evaluation capabilities and are
summarized in the following points:
\begin{list}{\labelitemi}{\leftmargin=1.5em \itemindent=-0.0em \itemsep=-.2em}
    \item Our primary contribution is a SpecGen evaluation framework that
        unifies multiple evaluation signals and organizes them by the questions
        they answer and the evaluation modality they require.
    \item We assemble a unified dataset for SpecGen evaluation, comprising
        350 existing Lean SpecGen tasks from the VERINA
        \cite{Ye:25:AI4Math:Verina} and CLEVER \cite{Thakur:26:NIPS:Clever}
        datasets, preserving source provenance information.
        We also include the VeriCoding \cite{Bursuc:25:preprint:Vericoding} and
        AlgoVeri \cite{Zhao:26:preprint:Algoveri} datasets, although these aren't
        used for the evaluations in this paper.
        Importantly, we do not claim to introduce newly authored datasets of
        SpecGen problems.
    \item Several of our behavioural metrics rely on our hand-engineered extraction
        tools which extract preconditions, postconditions, Lean signatures and
        declarations from the generated outputs.  Also, we provide compatible
        adapters for the VERINA and CLEVER evaluation harnesses to enable their
        use on the unified dataset.
    \item We present initial results covering 4 SpecGen configurations on the
        189 VERINA and 161 CLEVER tasks. Our authored controls separately
        demonstrate why accepting valid outputs, rejecting invalid outputs, and
        retaining required inputs deserve distinct feedback.
\end{list}

The strongest controlled finding is that perfect postcondition-test scores can
coexist with zero required-input recall. This demonstrates a blind spot of
postcondition-only feedback, not the prevalence of that error in model outputs.
We use compatibility checks and evidence-status reporting to distinguish
observed behaviour from unavailable measurements. These engineering checks
support the interpretation of the metrics rather than constituting the central
research claim.


%% file: sections/design.tex
\subsection{Evaluation-system Overview}

\input{figures/evaluation_system_fullwidth}

Our SpecGen evaluation system is summarized in Figure~\ref{fig:evaluation-system}.
It treats each generated specification as an artifact to be examined through
several complementary forms of evidence.
For every task, we combine the candidate with: (i) the available evaluation
context, i.e., the Lean environment; (ii) the reference specification available
in the dataset; (iii) the natural-language statement (user's intent); and, (iv)
optionally the labeled examples.
After representation-specific parsing and projection, independent evaluators
measure formal validity, reference similarity, reference equivalence, and
behavioral adequacy as detailed in Section~\ref{sec:metrics}.
Each result is stored together with its scope, denominator, runtime, and
evidence status.
Consequently, unsupported representations, missing dependencies, timeouts, and
evaluator failures are made distinguishable from observed metric failures, and
downstream reports can compare systems without treating unavailable evidence as
a score of zero.

\subsection{Datasets and SpecGen agents} \label{sec:agents}
We evaluate four productive SpecGen agent configurations on two datasets:
189 VERINA tasks \cite{Ye:25:AI4Math:Verina} and 161 CLEVER tasks
\cite{Thakur:26:NIPS:Clever}.
The evaluated configurations are:
\begin{list}{\labelitemi}{\leftmargin=1.5em \itemindent=-0.0em \itemsep=.1em}
    \item a direct-prompted LLM baseline, with no retries or post-processing;
    \item the VERINA prompting and rendering pipeline
        \cite{Ye:25:AI4Math:Verina};
    \item the CLEVER few-shot pipeline
        \cite{Thakur:26:NIPS:Clever}; and
    \item a LeanAide adapter \cite{Gadgil:22:NIPS:LeanAide}.
\end{list}
All configurations use the same base agentic LLM, GPT-5.6 (Sol), in our
experiments.
They mainly differ in their prompts, output representations, retry
budgets, and post-processing, thus representing a progression toward
increasingly structured SpecGen pipelines drawn from the literature.

\subsection{Supported SpecGen Evaluation Metrics} \label{sec:metrics}
In this section we describe the evaluation metrics used in our study.
Let $\widehat{Q}$ denote a generated postcondition candidate, $Q$ its reference
specification from the dataset, and $\mathcal{E}$ the recorded Lean environment.
We group the evaluation signals into four complementary metric families.
Binary and graded metrics are defined only when their required artifacts and
evaluator dependencies are available.

\begin{list}{\labelitemi}{\leftmargin=1.5em \itemindent=-0.0em \itemsep=.1em}
    \item \textbf{Formal validity:}
        This family asks whether the generated artifact is accepted as a Lean
        declaration in its intended environment.
        Following standard proof-assistant evaluation
        \cite{Moura:21:ICAD:Lean4}, the type-checking metric is a boolean signal
        indicating whether the compiler accepts the specification.
        This can catch missing dependencies and unsupported representations,
        contingent on timeouts, and evaluator failures by the Lean environment
        $\mathcal{E}$.
        It does not establish agreement with either the reference or the
        natural-language requirement.

    \item \textbf{Reference similarity:}
        This family compares $\widehat{Q}$ with $Q$ at textual, symbolic, and
        structural levels.
        Its metrics quantify resemblance to a supplied reference, but cannot
        establish logical equivalence rigorously.
        \begin{list}{\labelitemi}{\leftmargin=1.5em \itemindent=-0.0em \itemsep=0.0em}
            \item \ul{\emph{Normalized exact match.}}
                Let a normalization function $\mathcal{N}(\cdot)$ remove Lean4
                comments, potential Markdown in LLM generated outputs,
                and the first declaration name, and collapse whitespace.
                Then the normalized exact match (NEM) metric is a boolean signal
                indicating:
                \begin{equation}
                    \operatorname{NEM}(Q,\widehat{Q}) = \mathbf{1}\!\left\{
                        \mathcal{N}(\widehat{Q}) = \mathcal{N}(Q)
                    \right\},
                \end{equation}
                where $\mathbf{1}\{\cdot\}$ is the indicator function.  This is
                a project-specific normalization, so its definition, rather
                than an external source, determines the metric.

            \item \ul{\emph{Signature F1.}}
                Let $S_{\widehat{Q}}$ and $S_Q$ be the multisets of normalized
                features extracted from $\widehat{Q}$ and $Q$.
                For multisets $A$ and $B$, define their overlap by
                \begin{equation}
                    I(A,B)=\sum_{u}\min\!\left\{m_A(u),m_B(u)\right\},
                    \label{eq:multiset-overlap}
                \end{equation}
                where \(m_A(u)\) is the multiplicity of \(u\) in \(A\).
                Less formally, $I(A,B)$ counts the number of features that
                appear in both multisets.
                We define the signature precision, recall, and F1 as
                \begin{equation}
                    P_{\mathrm{sig}} = \frac{I(S_{\widehat{Q}},S_Q)}{|S_{\widehat{Q}}|},
                    \qquad
                    R_{\mathrm{sig}} = \frac{I(S_{\widehat{Q}},S_Q)}{|S_Q|},
                    \qquad
                    F_{1,\mathrm{sig}} = \frac{2P_{\mathrm{sig}}R_{\mathrm{sig}}}
                        {P_{\mathrm{sig}} + R_{\mathrm{sig}}}.
                \end{equation}
                Thus repeated binders and types contribute according to their
                multiplicities. The evaluator assigns F1 of one when both feature
                multisets are empty, and zero when only one is empty or their
                overlap is zero. Empty extraction can reflect unsupported
                declaration forms, so a score of one need not indicate matching
                signatures.

            \item \ul{\emph{Logical-atom F1.}}
                Let $A_{\widehat{Q}}$ and $A_Q$ be the multisets of normalized
                identifiers and supported logical and arithmetic operators
                extracted from the two statements. The current tokenizer omits
                numeric literals and the Unicode operators $\leq$, $\geq$, and
                $\neq$, limiting sensitivity to changes in these features.
                Using the same multiset overlap in \eqref{eq:multiset-overlap},
                we define the logical-atom precision, recall, and F1 as:
                \begin{equation}
                P_{\mathrm{atom}}
                    =\frac{I(A_{\widehat{Q}},A_Q)}{|A_{\widehat{Q}}|},
                \qquad
                R_{\mathrm{atom}}
                    =\frac{I(A_{\widehat{Q}},A_Q)}{|A_Q|},
                \qquad
                F_{1,\mathrm{atom}}
                    =\frac{2P_{\mathrm{atom}}R_{\mathrm{atom}}}
                        {P_{\mathrm{atom}} + R_{\mathrm{atom}}}.
                \end{equation}
                Signature and logical-atom F1 instantiate the standard F-score
                with different feature extractors.

            \item \ul{\emph{Generalized tree-edit similarity (or distance) (GTED).}}
                GTED \cite{Liu:25:preprint:GTED} standardizes each statement
                using Lean elaboration and its \emph{language server protocol}
                (LSP) hover information, constructs operator trees
                $\mathcal{T}_{\widehat{Q}}$ and $\mathcal{T}_Q$, and computes
                their tree-edit distance
                $d_{\mathrm{TED}}(\mathcal{T}_{\widehat{Q}},\mathcal{T}_Q)$.
                Matching the released GTED implementation in
                \citet{Liu:25:preprint:GTED}, we report the normalized
                similarity
                \begin{equation}
                \operatorname{GTED}(\widehat{Q},Q) =
                    1 - \frac{d_{\mathrm{TED}}(\mathcal{T}_{\widehat{Q}},\mathcal{T}_Q)}
                    {\max\{|\mathcal{T}_{\widehat{Q}}|,|\mathcal{T}_Q|\}},
                \end{equation}
                where $|T|$ is the number of nodes in $T$.
                Insertions, deletions, and unequal-label substitutions have unit
                cost; equal-label substitutions have zero cost.
                Consequently, $\operatorname{GTED}(\widehat{Q},Q) \in [0,1]$,
                with $1$ indicating identical operator trees under the adopted
                construction.
        \end{list}

        For reference similarity metrics, a high score does not necessarily
        indicate that the candidate and reference are logically equivalent.

    \item \textbf{Reference equivalence:}
        This family seeks formal evidence that the candidate and reference
        express equivalent propositions.
        Let $\pi(\widehat{Q})$ and $\pi(Q)$ denote \emph{projected} candidate and
        reference statements.
        BEq+ \cite{Poiroux:25:EMNLP:BEq} attempts to prove the bidirectional
        implication:
        \begin{equation}
            \mathcal{E},\pi(\widehat{Q}) \vdash \pi(Q) \qquad\text{and}\qquad
            \mathcal{E},\pi(Q) \vdash \pi(\widehat{Q}).
        \end{equation}
        Writing $q_{\rightarrow}$ and $q_{\leftarrow}$ for the corresponding
        Lean-checked proof outcomes, its success criterion is
        \begin{equation}
        \operatorname{BEq}^{+}(\widehat{Q},Q) = q_{\rightarrow} \land q_{\leftarrow}.
        \end{equation}
        A successful check involves constructing a Lean kernel checked evidence
        of equivalence for the compared projections.
        While BEq is one of the strongest signals of semantic agreement, constructing
        equivalence proofs is a challenging task in itself, and a proof-search
        such as this often fails.
        Failure to find either proof is classified as inconclusive, since search
        failure does not establish non-equivalence.

    \item \textbf{Behavioral adequacy:}
        For VERINA tasks with compatible candidate representations, this family
        evaluates the candidate on finite labeled examples.
        Let the candidate contract consist of a precondition $\widehat{P}(x)$ and
        postcondition $\widehat{Q}(x,y)$.
        Additionally, let $T^+$ contain intended input-output pairs that should
        pass the specification, $T^-$ the unintended pairs, and
        $D_{\mathrm{req}}$ the supplied required inputs.

        \begin{list}{\labelitemi}{\leftmargin=1.5em \itemindent=-0.0em \itemsep=.1em}
            \item \emph{Positive-test recall} is given by:
                \begin{equation}
                    R_{+}(\widehat{Q}) = \frac{1}{|T^+|} \sum_{(x,y)\in T^+}
                        \mathbf{1}\!\left\{\widehat{Q}(x,y)\right\}.
                \end{equation}
                It measures the fraction of supplied intended behaviors admitted
                by the candidate postcondition.

            \item \emph{Negative-test rejection} is given by:
                \begin{equation}
                    R_{-}(\widehat{Q}) = \frac{1}{|T^-|} \sum_{(x,y)\in T^-}
                        \mathbf{1}\!\left\{\neg \widehat{Q}(x,y)\right\}.
                \end{equation}
                It measures the fraction of supplied unintended behaviors
                excluded by the postcondition.

            \item \emph{Behavioral balanced accuracy} gives the two test classes
                equal weight, i.e.:
                \begin{equation}
                    \operatorname{BAcc}(\widehat{Q}) =
                    \frac{R_{+}(\widehat{Q})+R_{-}(\widehat{Q})}{2}.
                \end{equation}
                This is the usual balanced-accuracy construction
                \cite{Brodersen:10:ICPR:BAcc}, applied to behavioral acceptance
                and rejection.

            \item \emph{VERINA pass@1} is the binary finite-suite criterion
                given by:
                \begin{equation}
                    \operatorname{Pass@1}_{\mathrm{VERINA}}(\widehat{Q}) =
                        \mathbf{1}\!\left\{ R_{+}(\widehat{Q})=1 \land R_{-}(\widehat{Q})=1 \right\}.
                \end{equation}
                At the system level, the reported rate over tasks
                \(i=1,\ldots,n\) is
                \begin{equation}
                    \operatorname{Pass@1} = \frac{1}{n} \sum_{i=1}^{n}
                    \operatorname{Pass@1}_{\mathrm{VERINA}}(\widehat{Q}_i).
                \end{equation}
                This is the $k=1$ specialization of pass@$k$ reporting
                \cite{Chen:21:preprint:Codex}, with success defined here as
                passing both finite VERINA test classes.

            \item \emph{Required-input recall} evaluates the precondition
                independently:
                \begin{equation}
                    R_{\mathrm{req}}(\widehat{P}) = \frac{1}{|D_{\mathrm{req}}|}
                    \sum_{x\in D_{\mathrm{req}}}
                    \mathbf{1}\!\left\{\widehat{P}(x)\right\}.
                \end{equation}
                This prevents input domain coverage from being conflated with
                postcondition behavior.
                In particular, a false precondition can coexist with
                $R_{+} = R_{-} = 1$ while yielding $R_{\mathrm{req}} = 0$.
        \end{list}
        These scores establish behavior only on the supplied finite examples.
        They neither prove universal correctness nor show that the test oracle
        fully captures the natural-language requirement.

        \textbf{Soundness and completeness status:}
            Categorical summaries derived from the negative- and
            positive-example outcomes used above.
            They are not independent measurements, but rather a convenient way
            to report the two test-class results.
            We retain the benchmark's terminology while reporting $R_{+}$ and
            $R_{-}$ explicitly to make the operational meaning unambiguous.
\end{list}

The benchmark additionally provides \emph{LeanScorer}
\cite{Xuejun:25:preprint:LeanScorer}, a graded, requirement-conditioned LLM
assessment of intent alignment.
The integrated implementation aggregates its judgments using a \emph{Sugeno
fuzzy integral} \cite{Sugeno:72:SICE:Fuzzy}.
We treat it as an auxiliary learned assessment outside the four primary metric
families.

Evaluation time is recorded separately as telemetry.
Every conditional metric is reported with its applicable denominator and
evidence status, and unavailable evidence is represented by $\bot$, rather than
by a numeric zero or a semantic-failure label.

%% file: figures/evaluation_system_fullwidth.tex
\begin{figure}[t]
\centering
\resizebox{\textwidth}{!}{%
\begin{tikzpicture}[
  font=\sffamily\scriptsize,
  >=Latex,
  flow/.style={->,line width=.55pt,draw=black!60},
  box/.style={draw=black!45,rounded corners=1.5pt,align=center,inner sep=3pt},
  agentcontext/.style={box,fill=black!3,text width=38mm,minimum height=10mm},
  evalcontext/.style={box,fill=black!3,text width=42mm,minimum height=10mm},
  agent/.style={box,draw=NavyBlue!65,fill=NavyBlue!6,text width=22mm,minimum height=10mm},
  process/.style={box,draw=NavyBlue!65,fill=NavyBlue!6,text width=32mm,minimum height=10mm},
  metric/.style={box,draw=ForestGreen!65!black,fill=ForestGreen!7,text width=34mm,minimum height=12mm},
  result/.style={box,draw=BrickRed!65,fill=BrickRed!6,text width=57mm,minimum height=9mm}
]
\node[agentcontext] (agentctx) {\textbf{Agent context}\\Requirement and permitted prompt inputs};
\node[agent,right=3mm of agentctx] (agent) {\textbf{SpecGen}\\agent};
\node[process,right=3mm of agent] (candidate)
  {\textbf{Generated candidate}\\Parsing, normalization, projection};
\node[evalcontext,right=6mm of candidate] (evalctx)
  {\textbf{Evaluation-only context}\\Reference, labeled examples, Lean environment};

\node[metric,below=9mm of agentctx.south west,anchor=north west] (valid)
  {\textbf{Formal validity}\\Type-checking};
\node[metric,right=3mm of valid] (sim)
  {\textbf{Reference similarity}\\Exact match, F1, GTED};
\node[metric,right=3mm of sim] (equiv)
  {\textbf{Reference equivalence}\\BEq+};
\node[metric,right=3mm of equiv] (behav)
  {\textbf{Behavioral adequacy}\\Positive, negative, input tests};

\node[draw=ForestGreen!45!black,rounded corners=2pt,dashed,
      fit=(valid)(sim)(equiv)(behav),inner sep=2mm,
      label={[font=\sffamily\scriptsize\bfseries,ForestGreen!45!black]above:
      Complementary evaluators}] (family) {};

\node[result,below left=6mm and -19mm of family.south] (record)
  {\textbf{Evidence record}\\value/status, scope, denominator, runtime};
\node[result,right=6mm of record] (report)
  {\textbf{Coverage-aware reporting}\\aggregation and diagnostic comparison};

\draw[flow] (agentctx.east) -- (agent.west);
\draw[flow] (agent.east) -- (candidate.west);
\draw[flow] (candidate.south) -- ++(0,-2.5mm) -| ($(family.north west)!0.35!(family.north east)$);
\draw[flow] (evalctx.south) -- ++(0,-2.5mm) -| ($(family.north west)!0.78!(family.north east)$);
\draw[flow] (family.south) -- ++(0,-2mm) -| (record.north);
\draw[flow] (record) -- (report);
\end{tikzpicture}%
}

\caption{\label{fig:evaluation-system}Evaluation-system overview. The
SpecGen agent receives only its permitted generation context. The resulting
candidate is combined with evaluation-only artifacts by four complementary
evaluator families. Each evaluator emits a scoped result before coverage-aware
aggregation, so unavailable evidence remains distinct from metric failure.}

\end{figure}
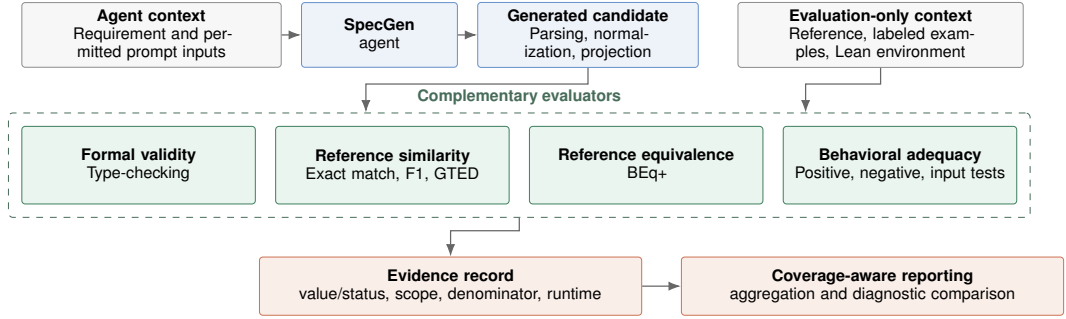

%% file: sections/results.tex
\subsection{Implementation details}

We analyze the performance of our four agents on VERINA and CLEVER datasets,
using the configurations introduced in Section~\ref{sec:agents}.
Each configuration attempts every task in the corresponding dataset, giving
756 VERINA and 644 CLEVER task--configuration records.
Generations are checkpointed, and their manifests record the model, prompt and
adapter fingerprints, and task identities.
Both runs configure medium reasoning effort for GPT-5.6 (Sol), an initial
output-token limit of 4,096, and a ceiling of 16,384;
however, due to varying prompts and harnesses, this cannot ensure equal realized
token use.
Because candidate budgets and repair procedures differ, these experiments
compare complete agent configurations, and they do not completely isolate the
effect of prompting or repair.

We report candidate production and recorded type-checking success over all
attempted tasks, including generation failures in the denominator.
Logical-atom F1 averages include the saved zero scores for generation failures.
GTED averages use available measurements, with their counts reported explicitly.
The shared reference projection uses Lean's Mathlib 4.20 library.
Our reference audit (Appendix~\ref{app:audits}) also checks the native VERINA
and CLEVER environments (Lean 4.18 and 4.24) to ensure consistency.
Behavioral results are available only for VERINA in these runs. CLEVER contains
test cases, but these runs lack the compatible VERINA-format test metadata
required by the behavioral evaluator.

\subsection{VERINA and CLEVER results}
\label{sec:experimental-findings}

\begin{figure}[t]
\centering
\includegraphics[width=\linewidth]{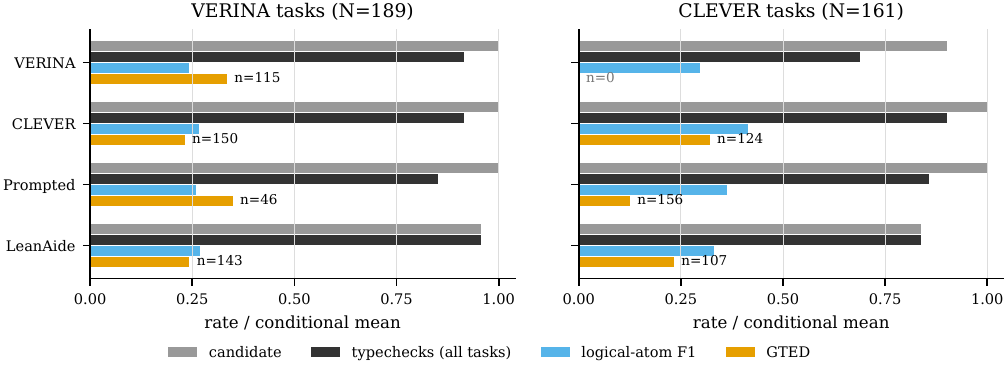}
\caption{Saved-run profiles on VERINA and CLEVER. Candidate production and
type-checking rates use all attempted tasks. Logical-atom F1 means include
saved generation-failure zeros. GTED means use available scores; annotations
give GTED sample sizes.
The VERINA configuration has no GTED measurements on CLEVER (\(n=0\)), not a
measured similarity of zero. Type-checking records can concern different
candidate projections or an adapter's internal check.}
\label{fig:agent-profiles}
\end{figure}

\paragraph{Formal validity and candidate coverage.}
Figure~\ref{fig:agent-profiles} shows that most configurations obtain high
recorded type-checking rates, but candidate availability also matters.
On VERINA, the VERINA and CLEVER configurations each return 189 candidates,
of which 173 (91.5\%) typecheck; the prompted baseline returns 189, with
161 (85.2\%) typechecking. LeanAide returns 181 candidates, all recorded as
typechecking, for an overall rate of 95.8\%. On CLEVER, the CLEVER configuration
and prompted baseline return all 161 candidates, with 145 (90.1\%) and
138 (85.7\%) typechecking. LeanAide's conditional rate remains 100\%, but
it returns only 135/161 candidates (83.9\%). The VERINA configuration returns
145/161 candidates and records 111/161 type-checking successes (68.9\%).
Thus, a perfect conditional type-checking rate can conceal failures to produce
a candidate. These observations do not establish that repair causes better
specifications, since the configurations differ in more than their repair steps.

\paragraph{Reference resemblance depends on both dataset and measurement coverage.}
The prompted baseline has the largest conditional mean GTED on VERINA,
0.349, but only 46/189 measurements (24.3\%). The VERINA configuration obtains
0.335 on 115 tasks, while CLEVER and LeanAide obtain 0.232 and 0.243 on
150 and 143 tasks. On CLEVER, the CLEVER configuration has mean GTED 0.321
on 124/161 tasks, compared with 0.233 on 107 tasks for LeanAide and 0.125
on 156 tasks for the prompted baseline. No GTED score is available for the
VERINA configuration on this dataset. The apparent ordering therefore changes
across datasets, and the averages describe different measurable subsets.
They cannot be interpreted as a controlled semantic-quality ranking.

\begin{figure}[t]
\centering
\includegraphics[width=\linewidth]{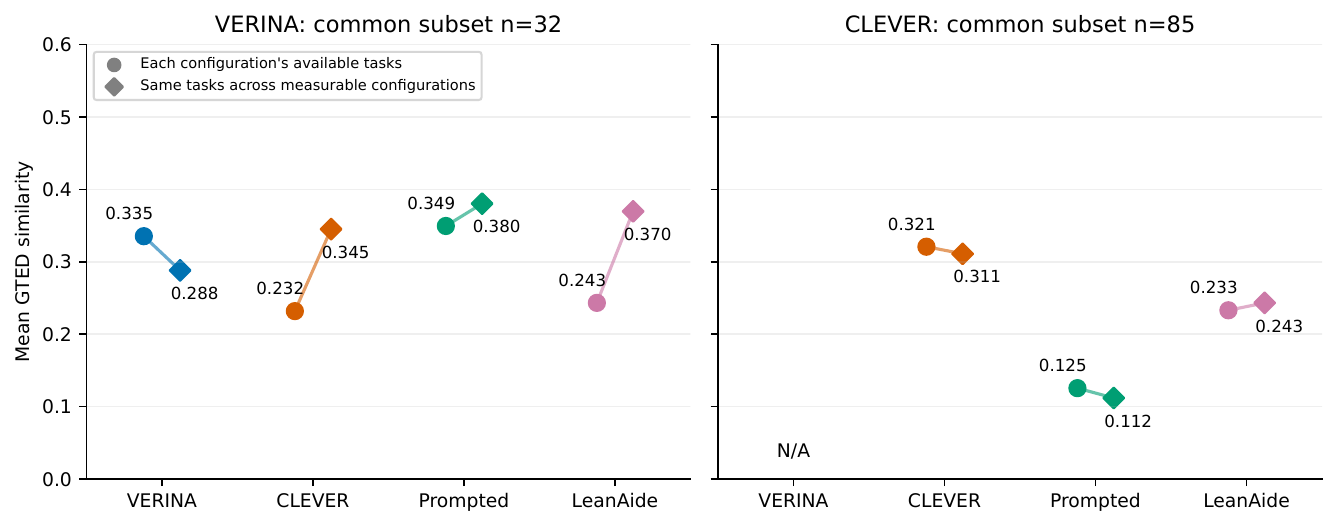}
\caption{GTED means before and after restricting to jointly measurable tasks.
Circles use each configuration's available measurements; diamonds use the
intersection of task IDs: 32 VERINA tasks across four configurations and
85 CLEVER tasks across three. The VERINA configuration has no GTED measurements
on CLEVER. Both comparisons retain the historical projections; matching tasks
does not establish semantic quality.}
\label{fig:gted-matched}
\end{figure}

The comparison in Figure~\ref{fig:gted-matched} uses each
configuration's available measurements for, say GTED.
This comparison is not controlled, since the support of each agent is different.
In order to normalize the comparison, we restrict the analysis to the
intersection of task IDs which have GTED measurements for all configurations.
This yields 32 tasks on VERINA and 85 tasks (excluding the VERINA agent) on
CLEVER.
The VERINA agent's mean decreases from 0.335 to 0.288,
while the CLEVER and LeanAide means increase from 0.232 and 0.243 to 0.345
and 0.370. The prompted baseline increases from 0.349 to 0.380. The VERINA
configuration therefore moves from second to fourth in observed mean GTED.
On CLEVER, the 85-task intersection among the three measurable configurations
preserves their ordering, with means of 0.311, 0.243, and 0.112 for CLEVER,
LeanAide, and the prompted baseline.
Thus, unequal measurement coverage can change the comparison, although it does
not do so uniformly across datasets.
Note that the intersections cover only 16.9\% and 52.8\% of the respective
datasets and are selected by metric availability; thus they are not
representative samples.

Logical-atom F1 adds a different view: on VERINA its means lie between
0.241 and 0.269 across configurations, whereas on CLEVER they range from
0.298 to 0.414. In particular, the prompted baseline's relatively high VERINA
GTED does not coincide with the highest logical-atom F1. Neither signal
supplies a correctness threshold. Moreover, inspecting the selected declaration
is necessary even when a structural score is available: the CLEVER examples in
Appendix~\ref{app:examples} show a reference helper selected in place of the
intended contract. These measurements characterize the recorded representations,
with that limitation, rather than establishing intent fidelity.
Appendix~\ref{app:gted-audit} examines the score distributions and concrete
normalization failures, including recorded GTED scores of one that compare
fragments rather than complete contracts.

\begin{figure}[t]
\change{\begin{minipage}{\linewidth}
\centering
\includegraphics[width=\linewidth]{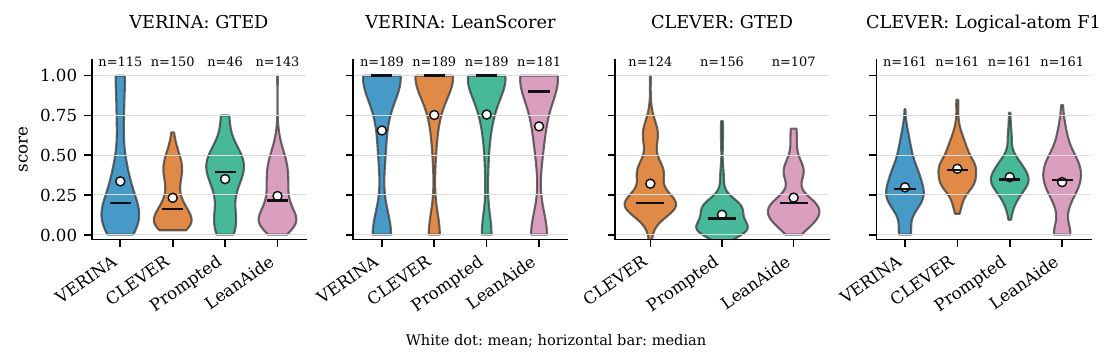}
\caption{Distributions of salient conditional metrics. White circles denote
means, horizontal bars denote medians, and annotations report available
measurements. Missing values are omitted rather than scored as zero.}
\label{fig:salient-metric-violins}
\end{minipage}}
\end{figure}

\change{\paragraph{Natural-language alignment.}
On the VERINA inventory, mean LeanScorer values are 0.655, 0.752, 0.755,
and 0.681 for VERINA, CLEVER, Prompted, and LeanAide, respectively, over
189, 189, 189, and 181 available generations. Figure~\ref{fig:salient-metric-violins}
shows the corresponding LeanScorer distributions alongside selected structural
metrics. Note that LeanScorer is an LLM-based judge and not a formal guarantee.
A complete scalar-metric table that tests the four agentic methods on VERINA and
CLEVER, is reported in Appendix~\ref{app:complete-metrics}.}

\paragraph{Behavioral outcomes require an evidence audit.}
Across the four VERINA configurations, the historical evaluator reports
185 finite-test passes, 469 nonpasses, and 102 unavailable records.
Of the 469 nonpasses, 357 (76.1\%) carry explicit harness errors;
112 lack sufficient evidence to identify a witnessed semantic failure.
For example, the VERINA configuration records 173 type-checking successes,
yet all 165 of its behavioral nonpasses have harness errors. Its attributed
precondition and postcondition declarations are not consistently preserved by
the behavioral projection. This disagreement primarily exposes incompatible
evaluation paths that prevent behavioral evaluation.

The CLEVER configuration, prompted baseline, and LeanAide have 95, 29, and
61 reported finite passes, respectively. These remain historical finite-test
outcomes, not independently revalidated correctness claims.
Six records have perfect positive-test recall but imperfect negative-test
rejection; unresolved proof or execution failures prevent counting them as
six confirmed intent errors. Appendix~\ref{app:audits} gives the complete
status breakdown. The CLEVER dataset has no behavioral measurements in these
runs, and neither dataset provides real-output required-input recall.
Consequently, the experiments do not estimate the prevalence of restrictive
preconditions in model-generated specifications.

\change{Although these runs cannot measure the prevalence of restrictive
preconditions, authored controls let us check whether the three behavioral
metrics respond separately to known postcondition and input-domain defects.
Appendix~\ref{app:controls} reports the controls, results, and full interpretation.}

%% file: sections/future.tex
\paragraph{Relationship to prior work.}
VERINA \cite{Ye:25:AI4Math:Verina} evaluates specification generation and
separates it from code and proof generation.
CLEVER's \cite{Thakur:26:NIPS:Clever} specification certification explicitly
targets agreement with a reference, making the trustworthiness and preservation
of that reference consequential.
LeanEuclid \cite{leaneuclid} supplies semantic evaluation in a controlled
geometry domain for mathematics; semantic evaluation itself is not our novelty.
VeriSpecGen \cite{verispecgen} goes further toward requirement-level validation
through traceability and targeted repair.
Our contribution is complementary: a unified evaluation framework relating
multiple signals to their feedback roles, with a controlled demonstration of why
input-domain coverage must be assessed separately from postcondition behaviour.
We explicitly identify the intent gap, which is a general phenomenon that can be
observed in multiple prior works \cite{vericontest}, and is a key challenge in
verifiable code generation.
We do not claim that prior literature ignored the intent gap, and our findings
merely provide evidence for this observation.
Importantly, we would also like to note that our evaluation framework was
implemented with the aid of Codex, which is an AI coding assistant, but was
rigorously tested and audited before evaluations were conducted.

\paragraph{Limitations.}
Our study covers one model and two Lean datasets, with configurations that
use different generation budgets and repair procedures. Uneven metric coverage,
projection failures, and historical evaluation errors limit comparisons of
specification quality. Reference agreement and finite tests do not establish
fidelity to the natural-language requirement. The authored controls demonstrate
a possible input-coverage failure, but its prevalence in generated outputs
remains unmeasured. We also do not establish that the metrics provide independent
information or improve generation when used as feedback.

\paragraph{Potential future directions.}
Future work should improve preservation of complete contracts across evaluators,
reassess affected scores, and measure required-input coverage in generated
specifications.
Independently reviewed references and checked counterexamples could strengthen
evaluation against user intent.
Studies across more models and domains, using matched budgets and held-out
tasks, should test whether combining these signals improves specification
quality over type-checking feedback alone, and at what generation cost.
A more comprehensive evaluation will extend to additional data sources in our
benchmarking suite, including VeriCoding \cite{Bursuc:25:preprint:Vericoding}
and AlgoVeri \cite{Zhao:26:preprint:Algoveri}.
Additionally, we plan to extend the system to incorporate Dafny
\cite{Leino:10:ICLPAIR:Dafny} and Verus \cite{Lattuada:23:ACM:Verus} SpecGen, as
well as principled evaluations of LEAN ProofGen.



%% file: sections/conclusion.tex
We presented a framework for evaluating specification generation through
multiple signals and applied it to existing VERINA and CLEVER tasks.
The results show why formal validity, reference resemblance, and finite
behavioral outcomes require separate interpretation: comparisons depend on
measurement coverage, and evaluator failures can masquerade as specification
failures. Authored controls further demonstrate that perfect postcondition
scores can coexist with exclusion of every required input. Together, these
findings support evaluating complete contracts and reporting the scope and
availability of each signal. Establishing whether this richer feedback
improves agents' fidelity to user intent remains an important next step.

%% file: sections/appendix_examples.tex
\section{Actual VERINA and CLEVER examples}
\label{app:examples}
These four rows come from saved benchmark runs, not authored controls. All
use the CLEVER configuration: the September 10 run on VERINA and the September
12 run on CLEVER. Problem statements are source excerpts; listings retain
all executable specification text, omitting only full-line comments. VERINA
references contain a precondition followed by a postcondition; CLEVER references
are complete declarations. Tables reproduce saved metrics rounded to three
decimals. TC denotes type-checking, NEM denotes normalized exact match, and
a dash denotes unavailable evidence. BEq 0 means no equivalence proof was
obtained, not a proof of inequivalence. Required-input recall is unavailable
for all four rows. Signature F1 is one in all four rows because both
extracted feature multisets are empty. This reflects the evaluator's
empty-feature convention and does not establish signature agreement.

\subsection{Integer square root: equivalence despite moderate similarity}
\label{app:example-sqrt}
\textbf{Dataset row:} VERINA, \texttt{verina\_basic\_65}.
\galleryproblem{This task involves computing the integer square root of a
given natural number. The goal is to determine the largest natural number
\(r\) that satisfies \(r*r\leq N\) and \(N<(r+1)*(r+1)\).}
\noindent\textbf{Generated specification}
\begin{lstlisting}[style=gallerylean]
def generated_spec
(impl: Nat → Nat)
(N: Nat) : Prop :=
let r := impl N
r * r ≤ N ∧ N < (r + 1) * (r + 1)
\end{lstlisting}
\noindent\textbf{Reference specification (precondition / postcondition)}
\begin{lstlisting}[style=gallerylean]
True

result * result ≤ N ∧ N < (result + 1) * (result + 1)
\end{lstlisting}
\input{data/example_gallery/sqrt_metrics}
\paragraph{Interpretation.}
The candidate binds \texttt{r := impl N} and imposes the same two bounds as
the reference. Substituting the observed result for the implementation output
makes these bounds coincide. The saved BEq check succeeds, and the behavioral
record reports all 5 positive and 13 negative examples passing, with no
harness-error flag. These remain recorded finite-test outcomes.
NEM is zero because the complete declaration and paired reference body are
textually different. Logical-atom F1 is 0.444 because
the feature comparison includes different declaration and binding forms,
not just the arithmetic relation. GTED compares an application of the
\texttt{generatedSpec} wrapper against the explicit reference formula. Its
recorded edit distance is 13 and larger tree size is 23, giving
\(1-13/23=0.435\). Moderate structural resemblance therefore coexists with
a successful equivalence check on this row.

\clearpage
\subsection{Sorted deduplication: behavioral zeros without counterexamples}
\label{app:example-dedup}
\textbf{Dataset row:} VERINA, \texttt{verina\_advanced\_81}.
\galleryproblem{Implement a Lean 4 function that, given a list of integers,
removes all duplicates and returns the resulting list in ascending order.}
\noindent\textbf{Generated specification}
\begin{lstlisting}[style=gallerylean]
def generated_spec
(impl: List Int → List Int)
(arr: List Int) : Prop :=
∃ result : List Int,
  impl arr = result ∧
  result.Nodup ∧
  result.Pairwise (fun a b : Int => a < b) ∧
  ∀ x : Int, x ∈ result ↔ x ∈ arr
\end{lstlisting}
\noindent\textbf{Reference specification (precondition / postcondition)}
\begin{lstlisting}[style=gallerylean]
True

List.isPerm arr.eraseDups result ∧ List.Pairwise (· ≤ ·) result
\end{lstlisting}
\input{data/example_gallery/dedup_metrics}
\paragraph{Interpretation.}
The reference requires a permutation of the deduplicated input and nondecreasing
order. The candidate states no duplicates, strict pairwise order, and equality
of input/output membership. These clauses describe the same intended sorted
set of integers through different predicates; the saved BEq search does not
establish their equivalence. Different predicate vocabulary also makes exact
match zero and logical-atom F1 low (0.171).
GTED retains a candidate wrapper application rather than expanding its clauses:
distance 15 over a larger tree of size 26 gives \(1-15/26=0.423\).

The historical behavioral evaluator records 0/5 positive successes and 0/15
negative rejections. Its diagnostics report failure to synthesize
\texttt{Decidable} for the existential and quantified test propositions.
For example, a negative test supplies input \([5,2,2,5]\) and output
\([2,2,5]\): the displayed candidate excludes that output through
\texttt{Nodup}, yet the attempted test fails to obtain decidability.
The recorded zeros do not demonstrate acceptance of this invalid output or
rejection of valid outputs. This row illustrates a successful type check
with unresolved behavioral evidence.

\clearpage
\subsection{Integer addition: an inconclusive equivalence search}
\label{app:example-add}
\textbf{Dataset row:} CLEVER, \texttt{add(x: Int, y: Int)} (row 52).
\galleryproblem{Add two numbers x and y.}
\noindent\textbf{Generated specification}
\begin{lstlisting}[style=gallerylean]
def generated_spec
(impl: Int → Int → Int)
(x y: Int) : Prop :=
impl x y = x + y
\end{lstlisting}
\noindent\textbf{Reference specification}
\begin{lstlisting}[style=gallerylean]
def problem_spec
(impl: Int → Int → Int)
(x y: Int) :=
let spec (res: Int) :=
  res - x - y = 0
∃ result, impl x y = result ∧
spec result
\end{lstlisting}
\input{data/example_gallery/add_metrics}
\paragraph{Interpretation.}
The candidate equates the implementation output with \(x+y\). The reference
introduces an existential result and requires \(\mathit{result}-x-y=0\).
Over integers, substituting the implementation output and rearranging the
equality gives the candidate's condition. This algebraic reading explains
why BEq 0 must remain inconclusive: it is not evidence that these contracts
disagree. The saved record provides no successful proof for this comparison.

Exact match is zero and logical-atom F1 is 0.538 because the reference uses subtraction,
an auxiliary predicate, and an existential equality rather than direct
addition. The GTED intermediates are \texttt{theorem TBD: := by sorry} and
\texttt{theorem spec : Int} \(\rightarrow\)
\texttt{Prop := by sorry}; distance 4 and larger tree size 5 yield
\(1-4/5=0.200\). These fragments do not preserve the displayed contracts,
so the score cannot be read as the similarity of their arithmetic meanings.
Behavioral fields are unavailable because this run lacks the VERINA-format
test metadata required by that evaluator, rather than because any test failed.

\clearpage
\subsection{Triangle area: reference agreement is a separate question}
\label{app:example-triangle}
\textbf{Dataset row:} CLEVER, \texttt{triangle\_area(a: float, h: float)}
(row 44).
\galleryproblem{Given length of a side and high return area for a triangle.}
\noindent\textbf{Generated specification}
\begin{lstlisting}[style=gallerylean]
def generated_spec
(impl: Rat → Rat → Rat)
(a h: Rat) : Prop :=
0 ≤ a ∧ 0 ≤ h → impl a h = (a * h) / 2
\end{lstlisting}
\noindent\textbf{Reference specification}
\begin{lstlisting}[style=gallerylean]
def problem_spec
(implementation: Rat → Rat -> Rat)
(a h: Rat) :=
let spec (result: Rat) :=
  a = 0 → result = 0 ∧
  (a ≠ 0 → (2 * result) / a = h);
∃ result, implementation a h = result ∧
spec result
\end{lstlisting}
\input{data/example_gallery/triangle_metrics}
\paragraph{Interpretation.}
For nonnegative base and height, the candidate requires the area \(a h/2\).
The reference's outer implication instead guards the entire conjunction
with \(a=0\). When \(a\neq0\), that implication places no constraint on the
result. Reading the displayed formulas at \(a=2,h=3\), an implementation
output of 999 satisfies the reference's local obligation but violates the
candidate's required output of 3. This is a direct logical reading of this
row, not a newly executed Lean check or a claim inferred from BEq 0. The
candidate also leaves behavior outside its nonnegative-input guard unspecified.

Both declarations typecheck. Logical-atom
F1 of 0.688 reflects shared arithmetic and comparison vocabulary, but does not
capture the consequence of moving an implication guard. NEM is zero. As in
the addition row, GTED normalization reduces the candidate to \texttt{TBD:}
and the reference to its local predicate type, giving distance 4, larger
tree size 5, and score 0.200. That score does not diagnose the guard issue.
No behavioral measurements are available. This example separates the
natural-language requirement, the generated contract, the reference contract,
and what the recorded metrics actually check.
\clearpage

%% file: data/example_gallery/sqrt_metrics.tex
\begin{center}\small
\begin{tabular}{cccccccc}
\toprule
TC & NEM & Sig. F1 & Atom F1 & GTED & BEq & \(R_+\) & \(R_-\)\\
\midrule
yes & 0 & 1.000 & 0.444 & 0.435 & 1 & 1 & 1\\
\bottomrule
\end{tabular}
\end{center}

%% file: data/example_gallery/dedup_metrics.tex
\begin{center}\small
\begin{tabular}{cccccccc}
\toprule
TC & NEM & Sig. F1 & Atom F1 & GTED & BEq & \(R_+\) & \(R_-\)\\
\midrule
yes & 0 & 1.000 & 0.171 & 0.423 & 0 & 0 & 0\\
\bottomrule
\end{tabular}
\end{center}

%% file: data/example_gallery/add_metrics.tex
\begin{center}\small
\begin{tabular}{cccccccc}
\toprule
TC & NEM & Sig. F1 & Atom F1 & GTED & BEq & \(R_+\) & \(R_-\)\\
\midrule
yes & 0 & 1.000 & 0.538 & 0.200 & 0 & -- & --\\
\bottomrule
\end{tabular}
\end{center}

%% file: data/example_gallery/triangle_metrics.tex
\begin{center}\small
\begin{tabular}{cccccccc}
\toprule
TC & NEM & Sig. F1 & Atom F1 & GTED & BEq & \(R_+\) & \(R_-\)\\
\midrule
yes & 0 & 1.000 & 0.688 & 0.200 & 0 & -- & --\\
\bottomrule
\end{tabular}
\end{center}

%% file: sections/appendix_audits.tex
\section{Audits supporting measurement validity}
\label{app:audits}

\paragraph{Source correspondence and environment compatibility.}
We distinguish full upstream contracts, exported contracts, and the shared
evaluator's projected statements. Native checks retain the required auxiliaries
and pre/post contracts while excluding implementations and correctness proofs.
All 350 upstream contracts compile in their pinned native environments.
Table~\ref{tab:refs} shows why a failed common projection cannot be attributed
directly to the source reference. The four CLEVER export failures use
\texttt{List.indexOf} where the upstream contracts use \texttt{List.idxOf};
this is a compatibility finding in Lean 4.24. The 16 CLEVER projection failures
involve missing import or source-local helper context. VERINA projection
failures also include declaration-selection and pre/post representation issues.
Because both context and representation change, this comparison does not
isolate a Lean-version effect.

\begin{table}[ht]
\centering\small
\caption{Reference compilation counts. Native and export checks use the source
environment; the common projection uses Mathlib 4.20 and a different artifact.}
\label{tab:refs}
\begin{tabular}{lrrrr}
\toprule
Dataset & Tasks & Upstream native & Export native & Common projection\\
\midrule
VERINA & 189 & 189 & 189 & 145\\
CLEVER & 161 & 161 & 157 & 145\\
\bottomrule
\end{tabular}
\end{table}

The source/export audit records 62 VERINA and 20 CLEVER textual differences.
Textual differences and an LLM-assisted alignment screen are not adjudicated
semantic defect counts. Likewise, native compilation does not certify the
reference's agreement with its natural-language requirement. The retained
counterexample recheck timed out importing Mathlib; we therefore do not use
that recheck as successful independent certification of reference witnesses.

\paragraph{Feature extraction and missing-score conventions.}
Inspection of the four examples in Appendix~\ref{app:examples} found empty
signature features on both sides of every comparison. Their saved Signature
F1 scores of one follow the empty-feature convention, without demonstrating
signature agreement. Logical-atom extraction omits numeric literals and some
Unicode comparison operators. We retain the historical scores and interpret
them within these limits. Aggregation also distinguishes saved logical-atom
zeros for generation failures from unavailable GTED measurements, which are
excluded from GTED means.

\subsection{Structural-score distributions and projection checks}
\label{app:gted-audit}
\begin{figure}[ht]
\centering
\includegraphics[width=\linewidth]{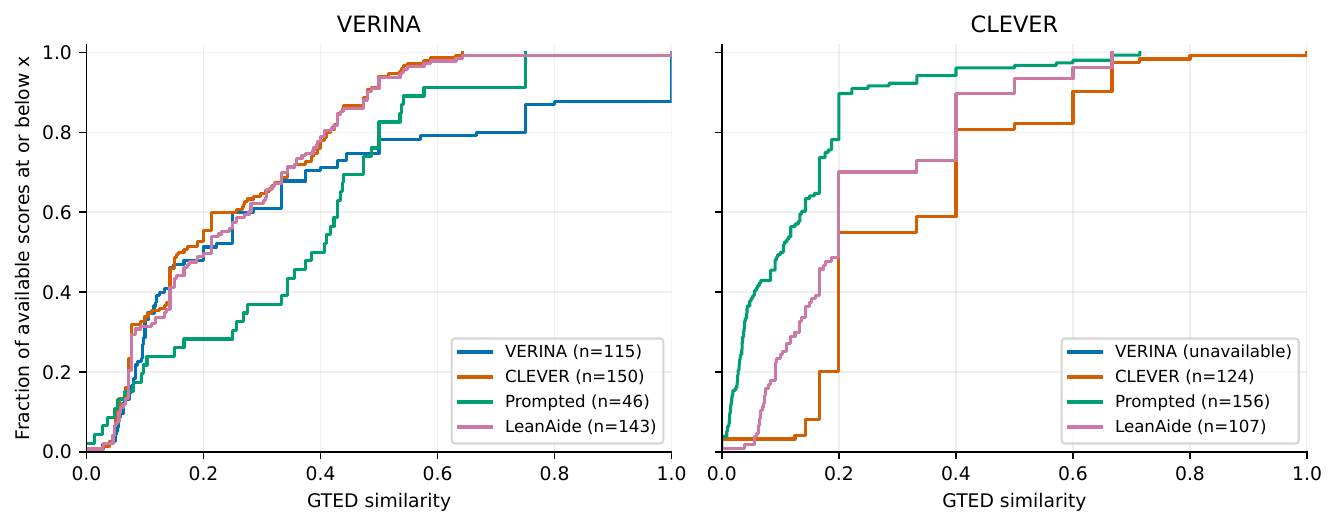}
\caption{Empirical cumulative distributions of available historical GTED
scores, with measurement counts in the legends. Each curve uses its own
available subset. These are distributions of recorded structural scores,
not semantic accuracy; inspected high-score examples contain incomplete
normalizations.}
\label{fig:gted-distributions}
\end{figure}

Figure~\ref{fig:gted-distributions} shows the distributions underlying the
conditional means. On VERINA, the VERINA configuration has mean GTED 0.335
but median 0.200; 14 of its 115 available scores equal one. The CLEVER,
prompted, and LeanAide configurations have medians 0.162, 0.396, and 0.214.
On CLEVER, the CLEVER and LeanAide configurations both have median 0.200,
despite means of 0.321 and 0.233; the prompted baseline has median 0.103
and mean 0.125. These distributions expose features hidden by averages,
but their upper tails require inspection of the compared artifacts.

For example, the VERINA configuration's output for
\texttt{verina\_basic\_39} has GTED 1 and logical-atom F1 0.182, but the
normalized comparison reduces to \texttt{l : List Int} versus
\texttt{result : List Int}. For \texttt{verina\_basic\_35}, GTED is also 1
and logical-atom F1 is 0.050, while the corresponding fragments are
\texttt{arr : Array Int} and \texttt{result : Array Int}.
For \texttt{verina\_advanced\_45}, GTED 1 coexists with logical-atom F1
0.082 and normalized theorem strings containing only \texttt{List} and
\texttt{TBD:}, respectively, before the placeholder proof.
These are evaluator intermediates, not the original generated specifications.
They show that a perfect recorded GTED score need not compare complete
contracts. We retain the historical scores and do not infer that all fourteen
perfect-score records share the same defect; they have not all been
individually adjudicated. Restricting comparisons to common task IDs in
Figure~\ref{fig:gted-matched} does not correct these normalization failures.

Within each dataset--configuration group with GTED measurements, Pearson
correlations between logical-atom F1 and GTED range from \(-0.238\) to
\(0.219\), computed on available pairs. The weak linear associations are
descriptive only: projection failures and discrete score patterns prevent
interpreting them as evidence of statistical independence or complementary
semantic information. Likewise, an inventory of numerically available
metrics is insufficient to establish valid evidence, because historical
behavioral scores can coexist with execution errors.

\subsection{Behavioral evidence and cohort checks}
\paragraph{Distinguishing failed evaluation from observed behavior.}
We inspect availability, explicit harness-error flags, and retained diagnostics
before interpreting a historical behavioral score. An explicit harness error
takes precedence over its number; unsupported cases and generation failures
remain separate. A failed proof attempt without a complete witness remains
unresolved. A reported finite pass requires both recorded test-class scores
to equal one without an inspected evaluator-error flag. Table~\ref{tab:agents}
preserves these distinctions for all 756 VERINA attempts.

\begin{table}[ht]
\centering\small
\caption{Retrospective VERINA evidence status, 189 attempts per configuration.
Pass denotes a reported finite pass; unresolved denotes a nonpass without
sufficient evidence of a semantic failure. No historical pass is upgraded by
this classification.}
\label{tab:agents}
\begin{tabular}{lrrrrr}
\toprule
Configuration & Pass & Harness error & Unresolved & Unsupported & No output\\
\midrule
VERINA & 0 & 165 & 0 & 24 & 0\\
CLEVER & 95 & 11 & 59 & 24 & 0\\
Prompted & 29 & 127 & 9 & 24 & 0\\
LeanAide & 61 & 54 & 44 & 22 & 8\\
\bottomrule
\end{tabular}
\end{table}

\paragraph{Executable outcomes and completeness of observations.}
The revised behavioral path evaluates decidable propositions and records
indexed Boolean outcomes, instead of inferring failed cases from failed
\texttt{native\_decide} proof attempts. Compilation failure, missing
decidability, or incomplete output withholds the score. The behavioral path
also rejects candidate commands and admissions such as \texttt{sorry} and
\texttt{axiom}, protecting the interpretation of executed test outcomes.
The controls in
Appendix~\ref{app:controls} document the available runtime evidence and its
remaining gap. A separate replay selected five non-demonstration task IDs by
the lowest SHA256 hashes of a fixed seed followed by each ID, shared across
all four configurations. However, the retained replay contains no completed
task results and reports Lean-session timeouts in its acceptance checks.
It does not revalidate the historical cohort or justify replacing its scores.

\paragraph{Cohort identity and demonstration overlap.}
Saved manifests associate task identities with generation fingerprints and
result-file hashes, allowing the audit to track the same outputs across
evaluation paths. The reference audit pins upstream revisions separately from
the generation adapters. The generation interface separates requirement and
signature context from evaluation-only references and tests. A secondary
VERINA profile excludes the two recorded demonstrations, leaving 187 tasks
per configuration. Its pass/harness-error/unresolved/unsupported/no-output
counts are, respectively, \(0/163/0/24/0\) for VERINA,
\(94/10/59/24/0\) for CLEVER, \(29/125/9/24/0\) for the prompted baseline,
and \(61/52/44/22/8\) for LeanAide. The coverage limitation persists after
this exclusion. These checks support traceability and interpretation; they
do not establish a budget-matched, held-out semantic leaderboard.

\change{\subsection{Complete metric comparison and benchmark dashboard}
\label{app:complete-metrics}
The complete comparison below reports every scalar signal used in the paper.
Bracketed counts make conditional coverage explicit; in particular, LeanScorer
was run on the VERINA paper inventory but not on CLEVER.}

\input{tables/all_agent_metrics}

\begin{figure*}[ht]
\change{\begin{minipage}{\textwidth}
\centering
\includegraphics[width=\textwidth]{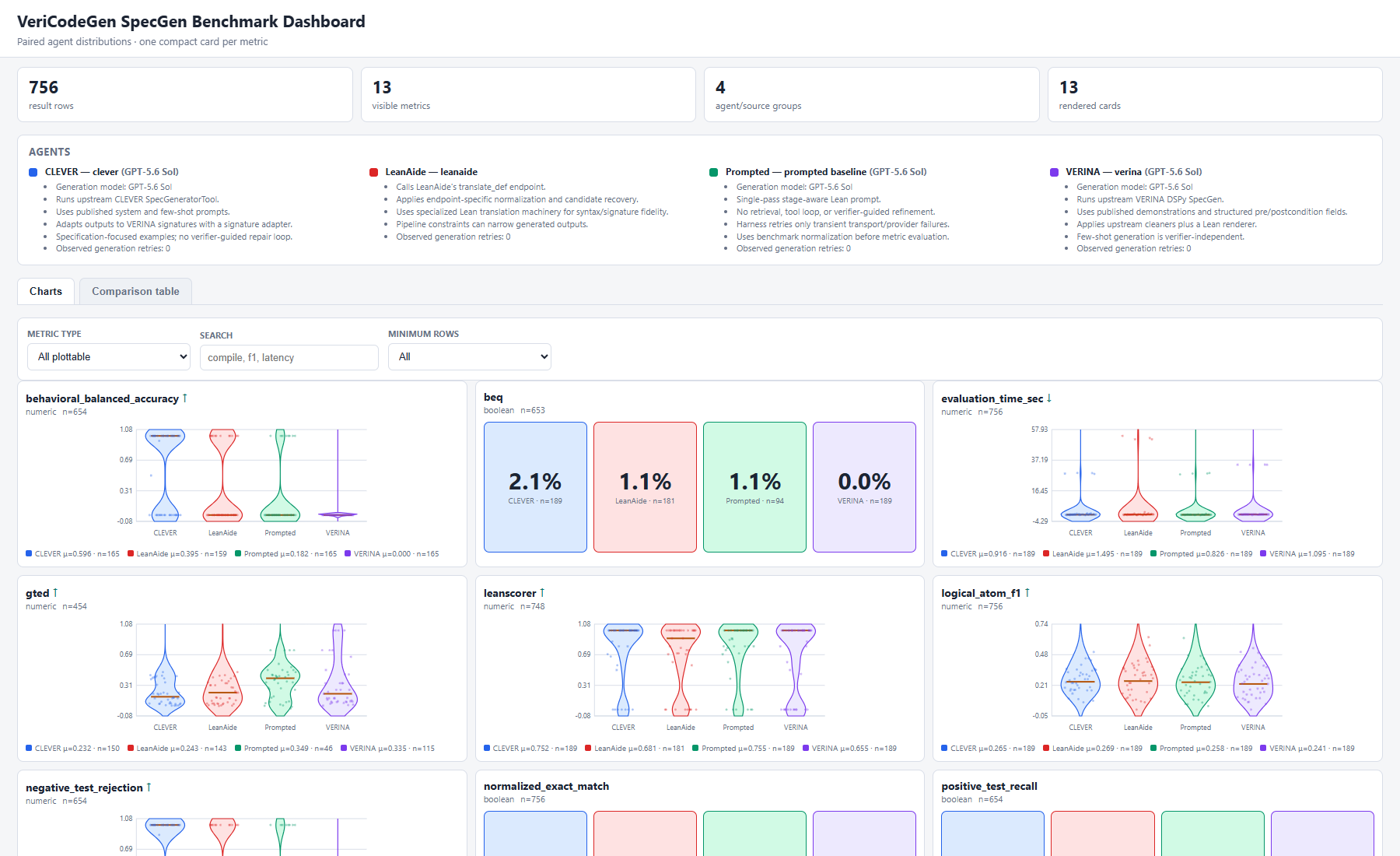}
\caption{Static dashboard generated by the benchmark suite for the 756 VERINA
task--agent records. The interface exposes metric availability, per-agent
distributions, summary values, and a comparison table.}
\label{fig:benchmark-dashboard}
\end{minipage}}
\end{figure*}

\subsection{Authored calibration controls}
\label{app:controls}
\begin{figure}[H]
\change{\begin{minipage}{\linewidth}
\centering
\includegraphics[width=.62\linewidth]{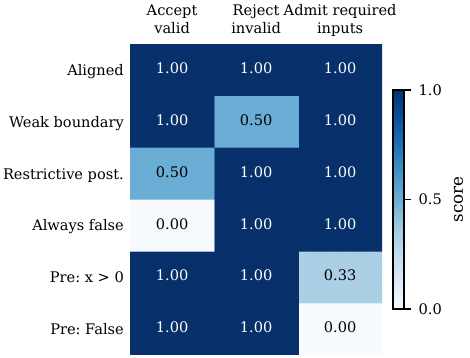}
\caption{Selected authored controls for the requirement to return an integer
\(y>x\). Positive recall and negative rejection each use six labeled pairs;
required-input recall uses the inputs \(-1,0,1\). The last two rows preserve
perfect postcondition scores while excluding required inputs. These controls
illustrate distinct feedback dimensions, not the prevalence of model errors.
This appendix gives all eight controls, including one with unavailable
postcondition measurements.}
\label{fig:controls}
\end{minipage}}
\end{figure}

\paragraph{Controlled changes to a relational specification.}
The authored requirement is: for any integer \(x\), return any integer
\(y>x\). Required inputs are \(-1,0,1\). For each input, the positive outputs
are \(x+1,x+2\), and the negative outputs are \(x,x-1\), giving six examples
per output class and three required-input checks. Table~\ref{tab:controls}
reports all eight saved controls. Figure~\ref{fig:controls} visualizes six
distinct measured patterns; the equivalent rewrite duplicates the aligned
scores, and the constant-true control has unavailable postcondition scores.

\begin{table}[ht]
\centering\small
\caption{Observed control scores. A dash denotes unavailable evidence.
The constant-true row is retained to expose the incomplete calibration.}
\label{tab:controls}
\begin{tabular}{llccc}
\toprule
Precondition & Postcondition & \(R_+\) & \(R_-\) & \(R_{\mathrm{req}}\)\\
\midrule
\(\mathrm{True}\) & \(y>x\) & 1 & 1 & 1\\
\(\mathrm{True}\) & \(x<y\) & 1 & 1 & 1\\
\(\mathrm{True}\) & \(y\geq x\) & 1 & \(1/2\) & 1\\
\(\mathrm{True}\) & \(y=x+1\) & \(1/2\) & 1 & 1\\
\(\mathrm{True}\) & \(\mathrm{True}\) & -- & -- & 1\\
\(\mathrm{True}\) & \(\mathrm{False}\) & 0 & 1 & 1\\
\(x>0\) & \(y>x\) & 1 & 1 & \(1/3\)\\
\(\mathrm{False}\) & \(y>x\) & 1 & 1 & 0\\
\bottomrule
\end{tabular}
\end{table}

The restrictive postcondition excludes \(x+2\), an allowed output under the
relational target. An implementation returning \(x+1\) would nevertheless be
a valid refinement of the requirement; the reduced recall measures lost
allowed behaviors, not an incorrect implementation. Conversely, the false
precondition excludes every required input while leaving the independently
tested postcondition unchanged. The constant-true control did not produce
postcondition measurements, so the retained run does not establish successful
calibration of every authored control.

\change{\paragraph{Controlled separation of the behavioral dimensions.}
In order to study the three behavioral dimensions, we authored eight
integer-return controls with known postconditions, preconditions, and required
inputs. Figure~\ref{fig:controls} provides a directly measured example of what
postcondition feedback omits. For the requirement to return an integer
\(y>x\), weakening the postcondition to \(y\geq x\) yields positive-test
recall 1 and negative-test rejection \(1/2\); restricting it to \(y=x+1\)
reverses these scores. Keeping \(y>x\) while changing the precondition to
\(x>0\) or \(\mathrm{False}\) preserves both postcondition scores at 1,
but reduces required-input recall to \(1/3\) or 0. These examples support
separate feedback for admitting valid outputs, rejecting invalid outputs,
and retaining required inputs; they do not measure the frequency of those
errors in generated outputs.}

%% file: tables/all_agent_metrics.tex
\begin{table*}[t]
\change{\begin{minipage}{\textwidth}
\centering
\caption{Mean metric values for the four evaluated agents. Bracketed values are the number of available measurements; candidate production and type-checking use all attempts. Best available values within each dataset are bold (ties included). LeanScorer was added later for the exact VERINA paper inventory and was not run on CLEVER.}
\label{tab:all-agent-metrics}
\small
\setlength{\tabcolsep}{5pt}
\begin{tabular}{lrrrr}
\toprule
Metric & VERINA & CLEVER & Prompted & LeanAide \\
\midrule
\multicolumn{5}{l}{\textbf{VERINA tasks}} \\
\addlinespace[2pt]
Candidate production & \textbf{1.000\,{\scriptsize [189]}} & \textbf{1.000\,{\scriptsize [189]}} & \textbf{1.000\,{\scriptsize [189]}} & 0.958\,{\scriptsize [189]} \\
Type-check rate (all attempts) & 0.915\,{\scriptsize [189]} & 0.915\,{\scriptsize [189]} & 0.852\,{\scriptsize [189]} & \textbf{0.958\,{\scriptsize [189]}} \\
Normalized exact match & 0.000\,{\scriptsize [189]} & 0.000\,{\scriptsize [189]} & 0.000\,{\scriptsize [189]} & 0.000\,{\scriptsize [189]} \\
Signature F1 & 0.117\,{\scriptsize [189]} & \textbf{0.836\,{\scriptsize [189]}} & 0.228\,{\scriptsize [189]} & 0.088\,{\scriptsize [189]} \\
Logical-atom F1 & 0.241\,{\scriptsize [189]} & 0.265\,{\scriptsize [189]} & 0.258\,{\scriptsize [189]} & \textbf{0.269\,{\scriptsize [189]}} \\
BEq+ & 0.000\,{\scriptsize [189]} & \textbf{0.021\,{\scriptsize [189]}} & 0.011\,{\scriptsize [94]} & 0.011\,{\scriptsize [181]} \\
GTED & 0.335\,{\scriptsize [115]} & 0.232\,{\scriptsize [150]} & \textbf{0.349\,{\scriptsize [46]}} & 0.243\,{\scriptsize [143]} \\
LeanScorer & 0.655\,{\scriptsize [189]} & 0.752\,{\scriptsize [189]} & \textbf{0.755\,{\scriptsize [189]}} & 0.681\,{\scriptsize [181]} \\
Behavioral balanced accuracy & 0.000\,{\scriptsize [165]} & \textbf{0.596\,{\scriptsize [165]}} & 0.182\,{\scriptsize [165]} & 0.395\,{\scriptsize [159]} \\
Positive-test recall & 0.000\,{\scriptsize [165]} & \textbf{0.594\,{\scriptsize [165]}} & 0.182\,{\scriptsize [165]} & 0.396\,{\scriptsize [159]} \\
Negative-test rejection & 0.000\,{\scriptsize [165]} & \textbf{0.598\,{\scriptsize [165]}} & 0.181\,{\scriptsize [165]} & 0.395\,{\scriptsize [159]} \\
VERINA pass@$1$ & 0.000\,{\scriptsize [165]} & \textbf{0.576\,{\scriptsize [165]}} & 0.176\,{\scriptsize [165]} & 0.384\,{\scriptsize [159]} \\
Evaluation time (s) & 1.0950\,{\scriptsize [189]} & 0.9159\,{\scriptsize [189]} & \textbf{0.8256\,{\scriptsize [189]}} & 1.4947\,{\scriptsize [189]} \\
\midrule
\multicolumn{5}{l}{\textbf{CLEVER tasks}} \\
\addlinespace[2pt]
Candidate production & 0.901\,{\scriptsize [161]} & \textbf{1.000\,{\scriptsize [161]}} & \textbf{1.000\,{\scriptsize [161]}} & 0.839\,{\scriptsize [161]} \\
Type-check rate (all attempts) & 0.689\,{\scriptsize [161]} & \textbf{0.901\,{\scriptsize [161]}} & 0.857\,{\scriptsize [161]} & 0.839\,{\scriptsize [161]} \\
Normalized exact match & 0.000\,{\scriptsize [161]} & 0.000\,{\scriptsize [161]} & 0.000\,{\scriptsize [161]} & 0.000\,{\scriptsize [161]} \\
Signature F1 & 0.012\,{\scriptsize [161]} & \textbf{1.000\,{\scriptsize [161]}} & 0.043\,{\scriptsize [161]} & 0.000\,{\scriptsize [161]} \\
Logical-atom F1 & 0.298\,{\scriptsize [161]} & \textbf{0.414\,{\scriptsize [161]}} & 0.362\,{\scriptsize [161]} & 0.331\,{\scriptsize [161]} \\
BEq+ & \textbf{0.000\,{\scriptsize [145]}} & \textbf{0.000\,{\scriptsize [161]}} & \textbf{0.000\,{\scriptsize [161]}} & \textbf{0.000\,{\scriptsize [135]}} \\
GTED & -- & \textbf{0.321\,{\scriptsize [124]}} & 0.125\,{\scriptsize [156]} & 0.233\,{\scriptsize [107]} \\
Evaluation time (s) & 0.0014\,{\scriptsize [161]} & 0.0013\,{\scriptsize [161]} & 0.0016\,{\scriptsize [161]} & \textbf{0.0007\,{\scriptsize [161]}} \\
\bottomrule
\end{tabular}
\end{minipage}}
\end{table*}